\documentclass[3p, 12pt]{elsarticle}\usepackage{graphicx, color}
\makeatletter
\def\maxwidth{ %
  \ifdim\Gin@nat@width>\linewidth
    \linewidth
  \else
    \Gin@nat@width
  \fi
}
\makeatother

\definecolor{fgcolor}{rgb}{0.2, 0.2, 0.2}
\newcommand{\hlfunctioncall}[1]{\textcolor[rgb]{0.501960784313725,0,0.329411764705882}{\textbf{#1}}}%
\newcommand{\hlstring}[1]{\textcolor[rgb]{0.6,0.6,1}{#1}}%

\usepackage{framed}
\makeatletter
\newenvironment{kframe}{%
 \def\at@end@of@kframe{}%
 \ifinner\ifhmode%
  \def\at@end@of@kframe{\end{minipage}}%
  \begin{minipage}{\columnwidth}%
 \fi\fi%
 \def\FrameCommand##1{\hskip\@totalleftmargin \hskip-\fboxsep
 \colorbox{shadecolor}{##1}\hskip-\fboxsep
     \hskip-\linewidth \hskip-\@totalleftmargin \hskip\columnwidth}%
 \MakeFramed {\advance\hsize-\width
   \@totalleftmargin\z@ \linewidth\hsize
   \@setminipage}}%
 {\par\unskip\endMakeFramed%
 \at@end@of@kframe}
\makeatother

\definecolor{shadecolor}{rgb}{.97, .97, .97}
\definecolor{messagecolor}{rgb}{0, 0, 0}
\definecolor{warningcolor}{rgb}{1, 0, 1}
\definecolor{errorcolor}{rgb}{1, 0, 0}
\newenvironment{knitrout}{}{} 

\usepackage{alltt}

\usepackage{texshade}
\usepackage{tikz}
\usetikzlibrary{shapes,arrows,shadows,fit}
\usetikzlibrary{positioning}
\usetikzlibrary{trees}
\pgfdeclarelayer{background}
\pgfdeclarelayer{foreground}
\pgfsetlayers{background,main,foreground}
\tikzstyle{fun} = [rectangle, draw, fill=white, 
  text width=5em, text centered, rounded corners, minimum height=2em]
\tikzstyle{obj} = [rectangle, draw, fill=gray!20, drop shadow, 
  text width=5em, text centered, rounded corners, minimum height=2em]

\newcommand{\R}{\texttt{R} }
\newcommand{\Rfunction}[1]{{\texttt{#1}}}
\newcommand{\Robject}[1]{{\texttt{#1}}}
\newcommand{\Rpackage}[1]{{\mbox{\normalfont\textsf{#1}}}}

\usepackage{hyperref}
\hypersetup{%
  pdfusetitle,
  pdfauthor={Laurent Gatto},
  bookmarks = {true},
  bookmarksnumbered = {true},
  bookmarksopen = {true},
  bookmarksopenlevel = 2,
  unicode = {true},
  breaklinks = {false},
  hyperindex = {true},
  colorlinks = {true},
  linktocpage = {true},
  plainpages = {false},
  linkcolor = {Blue},
  citecolor = {Blue},
  urlcolor = {Red},
  pdfstartview = {Fit},
  pdfpagemode = {UseOutlines},
  pdfview = {XYZ null null null}
}

\usepackage{amssymb}

\journal{BBA - Proteins and Proteomics}
\IfFileExists{upquote.sty}{\usepackage{upquote}}{}

\begin{document}

\begin{frontmatter}

\title{Using \R and Bioconductor for proteomics data analysis}

\author{Laurent Gatto\corref{cor1}\fnref{fn1}}
\ead{lg390@cam.ac.uk}
\cortext[cor1]{Corresponding author}
\fntext[fn1]{Tel: +44 1223 760253 ~ Fax: +44 1223 333345}

\author{Andy Christoforou}
\ead{ac587@cam.ac.uk}

\address{
  Cambridge Centre for Proteomics, 
  Department of Biochemistry\\ 
  University of Cambridge,
  Tennis Court Road, 
  Cambridge, CB2 1QR, UK}

\begin{abstract}
This review presents how \texttt{R}, the popular statistical environment and programming language, can be used in the frame of proteomics data analysis. A short introduction to \R is given, with special emphasis on some of the features that make \R and its add-on packages a premium software for sound and reproducible data analysis. The reader is also advised on how to find relevant \R software for proteomics. Several use cases are then presented, illustrating data input/output, quality control, quantitative proteomics and data analysis. Detailed code and additional links to extensive documentation are available in the freely available companion package \Rpackage{RforProteomics}.
\end{abstract}


\end{frontmatter}





\section{Introduction}\label{sec:intro}

Proteomics is evolving at a rapid pace \cite{Nilsson10} and updates in technologies and instruments applied to the study of bio-molecules, such as proteins or metabolites, require proper computational infrastructure \cite{Aebersold2011}. A broad diversity of complementary tools for data processing, management, visualisation and analysis have already been offered to the community and reviewed elsewhere \cite{Gonzalez-Galarza2012, Perez-Riverol}. The work presented here focuses on a particular type of software, namely \R \cite{R2012}, and the add-on \textit{packages} that enable extension in its functionality and scope, and their usefulness to the analysis of proteomics data.

\R is an open source statistical programming language and environment, originally created by Ross Ihaka and Robert Gentleman \cite{IhakaGentleman1996} at the University of Auckland and, since the mid-1997, developed and maintained by the R-core group. Originally utilised in an academic environment for statistical analysis, it is now widely used in public and private sector in a broad range of fields \cite{NYT2009}, including computational biology and bioinformatics. The success of \R can be attributed to several features including flexibility, a substantial collection of good statistical algorithms and high-quality numerical routines, the ability to easily model and handle data, numerous documentation, cross-platform compatibility, a well designed extension system and excellent visualisation capabilities to list some of the more obvious ones \cite{Gentleman2008}. These are some of the requirements that need to be fulfilled to tackle the complexity and high-dimensionality of modern biology. 

The focus of \R itself is and remains centred around statistics and data analysis. Functionality can however be extended through third-party packages, which bundle a coherent set of functions, documentation and data to address a specific problem and/or data type of interest. The Bioconductor project\footnote{\url{http://bioconductor.org/}} \cite{Gentleman2004}, initiated by Robert Gentleman, has a specific focus on computational biology and bioinformatics and represents a central repository for hundreds of software, data and annotation packages dedicated to the analysis and comprehension of high-throughput biological data, and promoting open source, coordinated, cooperative and open development of interoperable tools. The development and distribution of new packages is a very dynamic and important aspect of the \R software itself. Adherence to good development practice is crucial and enforced by the \R package development pipeline through a built-in checking mechanism, ensuring, among other things, proper package installation and loading, package structure, code validity and correct documentation. In addition, package development also provides multiple opportunities for unit and integration testing as well as reproducible research \cite{Gentleman2004a, Gentleman2005, Peng2009, Donoho2010, Peng2011} through the mechanism of literate programming \cite{Knuth1984} and \Rpackage{Sweave} \cite{Leisch2002} or \Rpackage{knitr} \cite{knitr} vignettes, which is crucially important from a scientific perspective.

Packages can be submitted to the main central repository, the Comprehensive \R Archive Network (CRAN) or to Bioconductor, which provides its own repository, to assure tighter software interoperation. In addition, any developer can easily set up private or public CRAN-style systems. Software management can become a tedious task when thousands of packages are distributed, many of which depend on each other and interoperate in complete pipelines. In \texttt{R}, this has been solved by providing dedicated package repositories as well as straightforward installation and updating mechanisms. 

Most importantly, \R and many packages are regarded as quality software \cite{Chambers2008}. They are aimed at users who want to explore and comprehend complex data for which there is often no predefined recipe. It is also a research tool to tackle new questions in innovative ways. The Bioconductor project, for example, has had a substantial impact on the field of microarrays through multi-disciplinary and cooperative method development and implementation, paving best practises for the current development of state-of-the-art high throughput genomics data analysis and comprehension. With respect to \texttt{R}'s contribution to other areas of bioinformatics and computational biology, it has also a lot to offer to proteomics. Biologists and proteomicists can gain immensely from autonomous data exploration and analysis. Bioinformaticians working in computational proteomics can use \R and specialised packages as an independent analysis and research framework or employ them to complement existing pipelines.

\bigskip

This manuscript presents a brief overview of some applications of the \R software to the analysis of MS-based quantitative proteomics data. We will review compliance of \R with open proteomics data standards, input/output capabilities, quantitation pipelines for label-free and labelled quantitation, quality control, quantitative data analysis and relevant annotation infrastructure. The review is accompanied by a package, \Rpackage{RforProteomics}, that provides the code to install a selection of relevant tools to reproduce and adapt the examples described below. Installation instruction are provided on the package's web page\footnote{\url{http://lgatto.github.com/RforProteomics/}}. Once installed, the package is loaded with the \Rfunction{library} function as shown below, to make its functionality available.

\begin{knitrout}\scriptsize
\definecolor{shadecolor}{rgb}{0.969, 0.969, 0.969}\color{fgcolor}\begin{kframe}
\begin{alltt}
> \hlfunctioncall{library}(\hlstring{"RforProteomics"})
\end{alltt}

{\ttfamily\noindent\itshape\color{messagecolor}{This is the 'RforProteomics' version 1.0.3.\\Run 'RforProteomics()' in R or visit \\'http://lgatto.github.com/RforProteomics/' to get started.}}\end{kframe}
\end{knitrout}

\section{Using \R in proteomics}\label{sec:user}

\subsection{Finding relevant software}\label{sec:find}

\R is a very dynamic \textit{ecosystem} \cite{Messerschmitt2003, Lungu2009} -- yearly \R and bi-annual Bioconductor releases, exponentially growing number of available packages \cite{Fox2009}, numerous active mailing lists and a community of hundreds of thousands of active users and developers in private and corporate environment \cite{NYT2009}. There are currently thousands of packages available through the official repositories, and new packages are published, discontinued or replaced by new, more elaborate alternatives on a daily basis. Providing an up-to-date and exhaustive list of packages is unachievable, even for a specified area of interest like proteomics, and would undoubtedly be out-dated too quickly to be useful. Dedicated pages are available however, that allow one to obtain an overview of some of the available packages in a specific area. CRAN maintains topic task views\footnote{\url{http://cran.r-project.org/web/views/}}, which are curated and maintained by experts. Each view provides a summary and some guidance on some of the growing number of CRAN packages that are useful for a certain topic. 
As of this writing, the Chemometrics and Computational Physics view features a total of 67 packages, some of which are dedicated to mass spectrometry and will be described later. 
The Bioconductor project provides a set of dedicated keywords to categorise packages, called \textit{biocViews}, that can be explored interactively\footnote{\url{http://www.bioconductor.org/packages/devel/BiocViews.html}}. For proteomics, most relevant candidates are \textit{MassSpectrometry} (in the \textit{Software/AssayTechnology} view with 21 packages) and \textit{Proteomics} (in the \textit{Software/BiologicalDomain} view, 35 packages), although numerous data analysis and annotation packages in other categories provide invaluable support, some of which will also be demonstrated below. 

\subsection{Getting suitable data}\label{sec:data}

Software development, evaluation and demonstration can not be envisioned without appropriate data. Although \R packages most often focus on software functionality, packages are also used to distribute experimental and annotation data, displayed in the \textit{AnnotationData} and \textit{ExperimentData} \textit{biocViews}. A specific \texttt{MassSpectrometryData} category, currently offering 5 packages, is dedicated for experimental data of interest here. Software packages often also distribute small data sets for illustration, demonstration and code testing. 

To exemplify some of the pipelines in this publication, we will make use of a larger, public data set, available from the ProteomeXchange\footnote{\url{http://www.proteomexchange.org/}} \cite{Hermjakob2006} ProteomeCentral repository (data \texttt{PXD000001}\footnote{Data DOI: \url{http://dx.doi.org/10.6019/PXD000001}}). 
In this TMT 6-plex \cite{Thompson2003} experiment, four exogenous proteins were spiked into an equimolar \textit{Erwinia carotovora} lysate with varying proportions in each channel of quantitation; yeast enolase (ENO) at 10:5:2.5:1:2.5:10, bovine serum albumin (BSA) at 1:2.5:5:10:5:1, rabbit glycogen phosphorylase (PHO) at 2:2:2:2:1:1 and bovin cytochrome C (CYT) at 1:1:1:1:1:2. Proteins were then digested, differentially labelled with TMT reagents, fractionated by reverse phase nanoflow UPLC (nanoACQUITY, Waters), and analysed on an LTQ Orbitrap Velos mass spectrometer (Thermo Scientific). 
Files in multiple format will be used to illustrate the input/output capabilities that are available to the proteomics audience. The companion package provides dedicated functions to directly download the data. 

\subsection{Proteomics standards and MS data input-output}\label{sec:stdio}

Proteomics is a very diverse field in terms of applications, experimental designs and file formats. When dealing with a wide range of data, flexibility is often key; this is particularly relevant for the \R environment, which can be used for many different purposes and data types. Raw mass spectrometry data comes in many different formats. While closed vendor-specific binary formats are less interesting due to their limited scope, several research groups as well as the HUPO Proteomics Standards Initiative (PSI) have developed open XML-based standards, formats and libraries to facilitate the development of vendor-agnostic tools and analysis pipeline. This functionality is available through the \Rpackage{mzR} package \cite{mzR2012, Chambers2012}, that provides a unified interface to the \texttt{mzData} \cite{Orchard2007}, \texttt{mzXML} \cite{Pedrioli2004}, \texttt{mzML} \cite{Martens2010} as well as \texttt{netCDF} formats. The \Rfunction{openMSfile} function opens a connection to any of these file types and enables to query instrument information and raw data in a consistent way. It is generally used by experienced users or developers who require maximal flexibility. For instance, \Rpackage{mzR} is used by \Rpackage{xcms} \cite{Smith2006, Benton2008}, \Rpackage{TargetSearch} \cite{TargetSearch2009} and \Rpackage{MSnbase} \cite{Gatto2012} for interaction with raw data.

Other packages provide higher level interfaces to raw data, modelled as computational data containers that store data and meta-data while assuring internal coherence. Such \textit{classes} come with a set of associated \textit{methods}, that allow the application of predefined actions on class instances, also called \textit{objects}, such as accessing specific pieces of information, modifying parts of the data or producing relevant graphical representation of the data.
The \Robject{MSnExp} or \Robject{xcmsRaw} classes, defined in the \Rpackage{MSnbase} and \Rpackage{xcms} packages respectively, represent experiments as a collection of annotated spectra, with the aim of removing the burden of users to manipulate the complex data by bundling it in specialised classes with an easy-to-use and well documented interface, the associated methods, to streamline the most common tasks. The example raw file used below, available from the \Rpackage{MSnbase} package, is an iTRAQ 4-plex \cite{Ross2004} experiment. It is read into \R and converted into an \Robject{MSnExp} object using the \Rfunction{readMSData} function. This specific data structure allows the spectra to be stored along with associated meta data and enables easy manipulation of the complete annotated data set. The last line displays a summary of the data in the \R console and figure \ref{fig:msnexp} illustrates some of the raw data plotting functionality applicable to an \Robject{MSnExp} instance (left) or an individual spectrum (right).

This first command finds the location of the test data file.

\begin{knitrout}\scriptsize
\definecolor{shadecolor}{rgb}{0.969, 0.969, 0.969}\color{fgcolor}\begin{kframe}
\begin{alltt}
> mzXML <- \hlfunctioncall{dir}(\hlfunctioncall{system.file}(package = \hlstring{"MSnbase"}, dir = \hlstring{"extdata"}), 
+     full.name = TRUE, pattern = \hlstring{"mzXML$"})
\end{alltt}
\end{kframe}
\end{knitrout}

We then proceed by reading the \texttt{mzXML} file and create an \Robject{MSnExp} object.

\begin{knitrout}\scriptsize
\definecolor{shadecolor}{rgb}{0.969, 0.969, 0.969}\color{fgcolor}\begin{kframe}
\begin{alltt}
> rawms <- \hlfunctioncall{readMSData}(mzXML, verbose = FALSE)
\end{alltt}
\end{kframe}
\end{knitrout}

Finally, we show a summary of the contents of the data object.

\begin{knitrout}\scriptsize
\definecolor{shadecolor}{rgb}{0.969, 0.969, 0.969}\color{fgcolor}\begin{kframe}
\begin{alltt}
> rawms
\end{alltt}
\begin{verbatim}
Object of class "MSnExp"
 Object size in memory: 0.2 Mb
- - - Spectra data - - -
 MS level(s): 2 
 Number of MS1 acquisitions: 1 
 Number of MSn scans: 5 
 Number of precursor ions: 5 
 4 unique MZs
 Precursor MZ's: 437.8 - 716.34 
 MSn M/Z range: 100 2017 
 MSn retention times: 25:1 - 25:2 minutes
- - - Processing information - - -
Data loaded: Tue Apr  9 22:10:44 2013 
 MSnbase version: 1.9.1 
- - - Meta data  - - -
phenoData
  rowNames: 1
  varLabels: sampleNames fileNumbers
  varMetadata: labelDescription
Loaded from:
  dummyiTRAQ.mzXML 
protocolData: none
featureData
  featureNames: X1.1 X2.1 ... X5.1 (5 total)
  fvarLabels: spectrum
  fvarMetadata: labelDescription
experimentData: use 'experimentData(object)'
\end{verbatim}
\end{kframe}
\end{knitrout}

\begin{figure}[!ht]
\begin{knitrout}\scriptsize
\definecolor{shadecolor}{rgb}{0.969, 0.969, 0.969}\color{fgcolor}

{\centering \includegraphics[width=.49\linewidth]{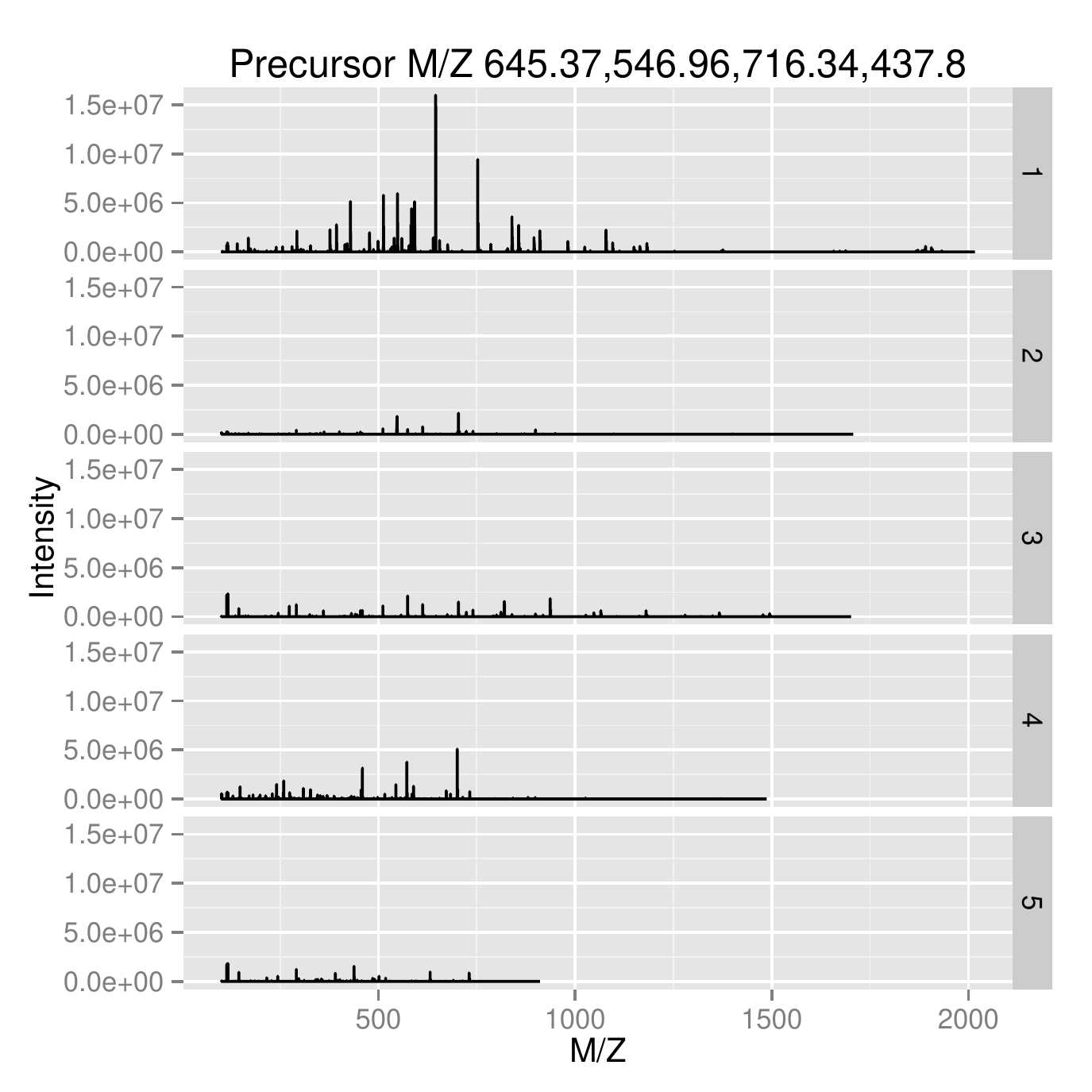} 
\includegraphics[width=.49\linewidth]{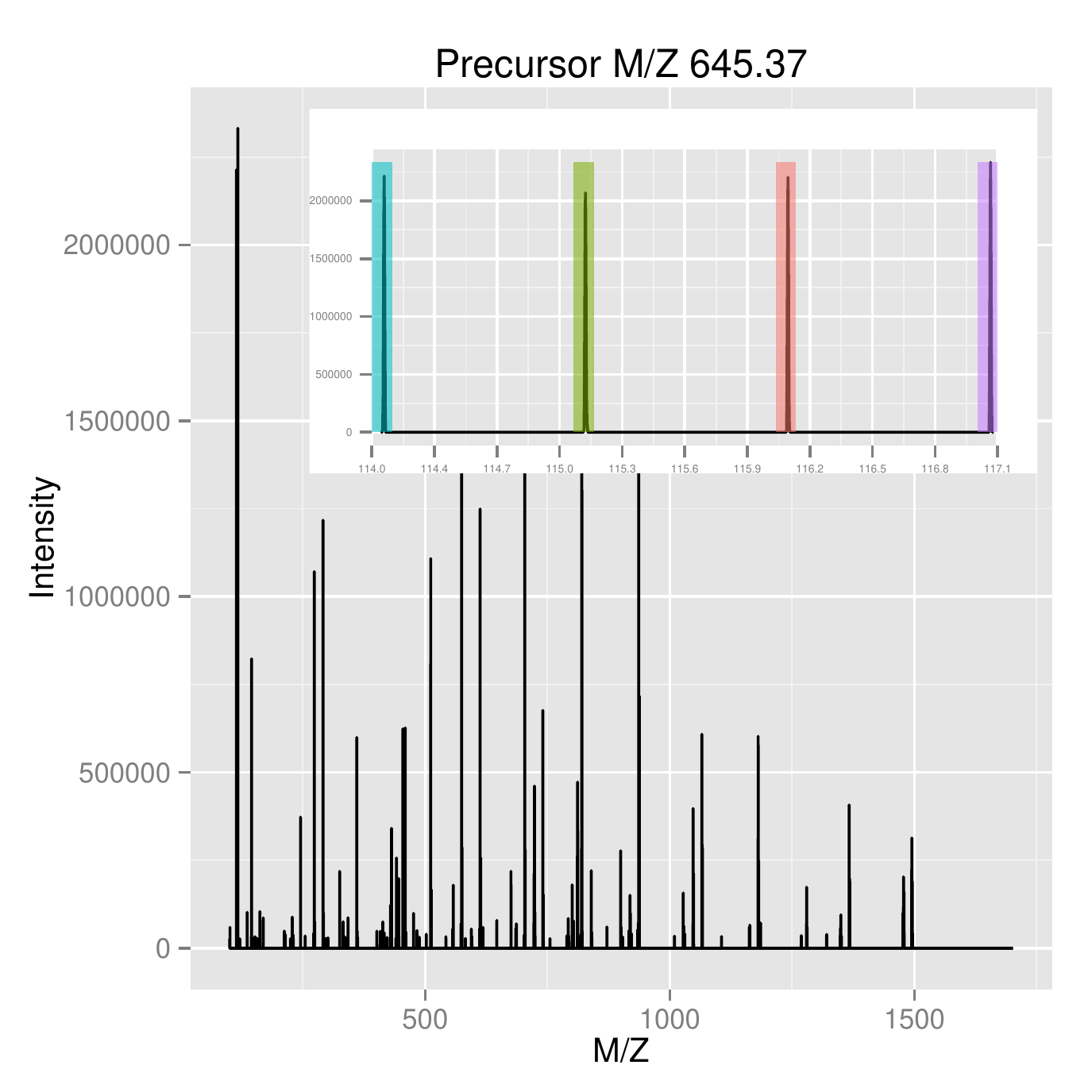} 

}

\end{knitrout}

\caption{Plotting raw MS$^2$ data using functionality from the \Rpackage{MSnbase} package. On the left, the full \textit{m/z} range of an experiment containing 5 spectra is displayed. On the right, one spectrum of interest is illustrated, highlighting the 4 iTRAQ reporter region. Both figures, have been created with the generic \Rfunction{plot} function, applied to either the complete experiment of a single MS$^2$ spectrum.}
\label{fig:msnexp}
\end{figure}

The \texttt{mgf} file format is also supported, for reading through the function \Rfunction{readMgfData}, which encapsulates the peak list data into \Robject{MSnExp} objects as above, and for writing such objects to a file through the \Rfunction{writeMgfData}. Other input/output facilities for quantified data will be presented in the next section. 

Standard formats for identification data are not yet systematically supported. It is however possible to import such information into \R, using existing \R data import/export infrastructure. For example, the \Rpackage{XML} package \cite{xml2012} allows one to parse arbitrary \texttt{xml} files based on their schema definition. Support for \texttt{mzIdentML}, \texttt{mzQuantML} and possible other community supported formats will be added to the \Rpackage{mzR} package.

\subsection{Data processing and quantitation}\label{sec:quant}

Quantitation has become an essential part of proteomics, and several alternatives are available in \R for label-free and labelled approaches. In this section, we will present quantitation functionality and associated raw data processing capabilities.

\subsubsection{Label-free quantitation}\label{ref:ms1quant}

Several packages provide functionality that can be applied to the analysis of label-free MS data. Although its first scope is the study of metabolites, \Rpackage{xcms} is a mature package that provides a complete pipeline for preprocessing LC/MS data for relative quantitation and data visualisation \cite{Mueller2008, Lange2008}. A typical \Rpackage{xcms} work flow implements peak extraction, filtering, retention time correction and matching across samples. The package is very versatile, featuring, for example, several peak picking methods, including some applying continuous wavelet transformation (CWT) \cite{Du2006, Tautenhahn2008}. The pipeline offers a complete framework to support data analysis and visualisation of chromatograms and peaks to be deemed to be differentially expressed. On-line help is available though a dedicated forum\footnote{\url{http://metabolomics-forum.com/}}. 

\begin{figure}[!htb]
  \begin{center}
    \includegraphics[width=1\linewidth]{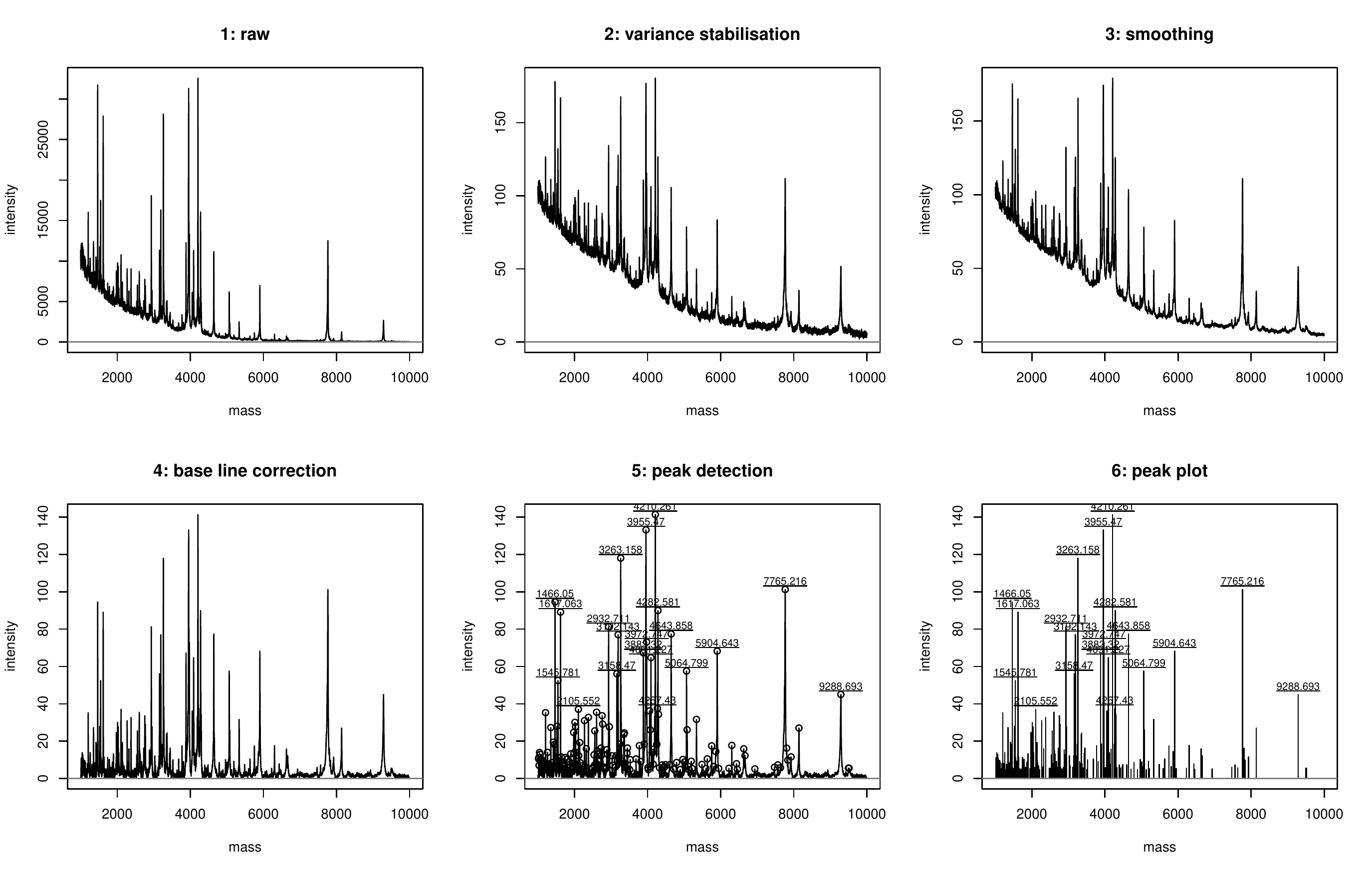}
    \caption{ Label-free spectrum processing peak detection from the \Rpackage{MALDIquant} package. Figures represent (1) raw data, (2) effect of variance stabilisation using square root transformation, (3) smoothing using a simple 5 point moving average, (4) base line correction, (5) noise reduction and peak detection and (6) final results. }
    \label{fig:maldi}        
  \end{center}
\end{figure}

\Rpackage{MALDIquant} \cite{Gibb2012} also provides a complete analysis pipeline for MALDI-TOF and other label-free MS data. Its distinctive features include baseline subtraction using the SNIP algorithm \cite{Ryan1998}, peak alignment using warping functions, handling of replicated measurements as well as supporting spectra with different resolutions. Figure \ref{fig:maldi} illustrates spectrum preprocessing and peak detection steps.

\Rpackage{synapter} is a package \cite{synapter2012} dedicated to the re-analysis of data independent MS$^E$ data \cite{Silva2006, Geromanos2009}, acquired on Waters Synapt instruments . It implements robust data filtering strategies, calculating and using peptide identification reliability statistics, peptide-to-protein ambiguity and mass accuracy. It then models retention time deviations between reliable sets of peptides in different runs and transfer identification across acquisitions to increase the overall peptide and protein coverage in full experiments through an easy-to-use interface. As illustrated in section \ref{sec:analysis}, it interoperates well with \Rpackage{MSnbase} to take advantage of the existing data structure and offers a complete analysis pipeline. 

Finally, packages that implement MS$^2$ data processing, like \Rpackage{MSnbase} and \Rpackage{isobar} \cite{Breitwieser2011} (see section \ref{sec:ms2quant}), also support spectral counting once identification data is available. In addition, \Rpackage{isobar} allows one to perform emPAI \cite{Ishihama2005} and distributed normalised spectral abundance factor (dNSAF) \cite{Zhang2010} quantitation.

\subsubsection{Labelled quantitation}\label{sec:ms2quant}

Pipelines for labelled MS$^2$ quantitation, using isobaric tagging reagents such as iTRAQ and TMT are available in the \Rpackage{isobar} and \Rpackage{MSnbase} packages. The code chunk below, taken from \Rpackage{MSnbase}, illustrates how to quantify the iTRAQ reporter peaks from the \Robject{rawms} data instance read in section \ref{sec:stdio}. The \Rfunction{quantify} function returns another data container, an \Robject{MSnSet}, specialised for storing quantitative data and associated meta data. Reporter impurity correction can then be applied using the \Rfunction{purityCorrect}. The \Rpackage{isobar} package imports centroided peak data identification data from \texttt{mgf} and text spread sheet files or converts \Robject{MSnSet} instances to create its own \Robject{IBSpectra} containers for further isotope impurity correction, normalisation and differential expression analysis (section \ref{sec:analysis}).

Below, we perform quantitation of the raw \Robject{MSnExp} data using the iTRAQ 4-plex reporters ions to create a new \Robject{MSnSet} object containing the quantitative data.

\begin{knitrout}\scriptsize
\definecolor{shadecolor}{rgb}{0.969, 0.969, 0.969}\color{fgcolor}\begin{kframe}
\begin{alltt}
> qnt <- \hlfunctioncall{quantify}(rawms, reporters = iTRAQ4, verbose = FALSE)
\end{alltt}
\end{kframe}
\end{knitrout}

In the following code chunk, we first define the reporter tag impurities as reporter by the manufacturer, apply the correction and display a summary of the resulting \Robject{MSnSet} instance.

\begin{knitrout}\scriptsize
\definecolor{shadecolor}{rgb}{0.969, 0.969, 0.969}\color{fgcolor}\begin{kframe}
\begin{alltt}
> impurities <- \hlfunctioncall{matrix}(\hlfunctioncall{c}(0.929, 0.059, 0.002, 0.000,
+                        0.020, 0.923, 0.056, 0.001,
+                        0.000, 0.030, 0.924, 0.045,
+                        0.000, 0.001, 0.040, 0.923),
+                      nrow=4)
> qnt <- \hlfunctioncall{purityCorrect}(qnt, impurities)
> qnt
\end{alltt}
\begin{verbatim}
MSnSet (storageMode: lockedEnvironment)
assayData: 5 features, 4 samples 
  element names: exprs 
protocolData: none
phenoData
  sampleNames: iTRAQ4.114 iTRAQ4.115 iTRAQ4.116
    iTRAQ4.117
  varLabels: mz reporters
  varMetadata: labelDescription
featureData
  featureNames: X1.1 X2.1 ... X5.1 (5 total)
  fvarLabels: spectrum file ... collision.energy (12
    total)
  fvarMetadata: labelDescription
experimentData: use 'experimentData(object)'
Annotation: No annotation 
- - - Processing information - - -
Data loaded: Tue Apr  9 22:10:44 2013 
iTRAQ4 quantification by trapezoidation: Tue Apr  9 22:10:49 2013 
Purity corrected: Wed May 15 14:32:10 2013 
 MSnbase version: 1.9.1 
\end{verbatim}
\end{kframe}
\end{knitrout}

\begin{figure}[!htb]
  \begin{center}
    \includegraphics[width=0.75\linewidth]{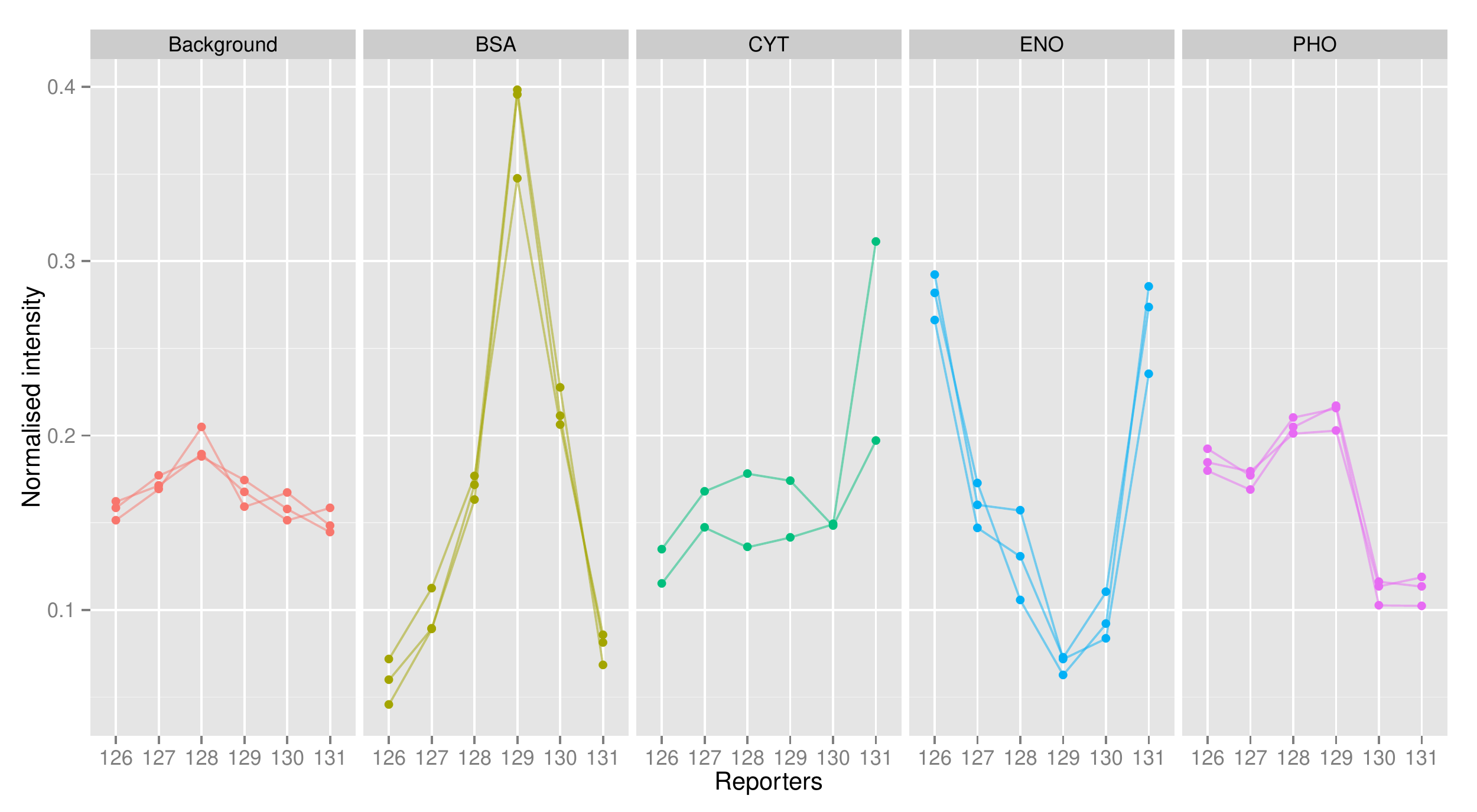}
    \caption{Representation of peptide-level quantitation data. This plot has been generated using the \texttt{PXD000001} TMT 6-plex data and converted to an \Robject{MSnSet} object. Normalised background and spike (BSA, CYT, ENO and PHO) reporter ion intensities for a subset of peptides have been plotted using the \Rpackage{ggplot2} package \cite{Wickham2009}. The complete code is available in the companion package.}
    \label{fig:spikes}
  \end{center}
\end{figure}

Once spectrum-level data is produced and stored in the specialised containers with peptide identification and protein inference meta data, it can be visualised (see figure \ref{fig:spikes}) and combined into peptide- and protein-level quantitation data.

Data analysis capabilities, including data normalisation and statistical procedures, are well known strengths of the \R software. It is therefore important to provide support for the exchange of quantitative data. The newly developed \texttt{mzTab}\footnote{\url{https://code.google.com/p/mztab/}} file, that aims at facilitating proteomics and metabolomics data dissemination to a wider audience through familiar spreadsheet-based format, can also be incorporated and exported using the \Rfunction{readMzTabData} and \Rfunction{writeMzTabData} functions. It is of course also possible to import quantitation data exported by third party applications to spread sheet formats. The most general way to import such data is using the \Rfunction{read.table} function. Specialised alternatives exist, to produce data structures, like \Robject{MSnSet}s. The \Rfunction{readMSnSet} function, for instance, can import quantitation data, feature meta data and sample annotation from spread sheets and create fully-fledged \Robject{MSnSet} instances.

\bigskip

Additional packages provide specialised functionalities relevant to  data processing. \Rpackage{IPPD} \cite{IPPD2012}  uses template matching to deconvolute peak patterns in individual raw spectra or complete experiments. \Rpackage{Rdisop} \cite{Bocker2008, Bocker2009} is designed to determine the formula of ions based on their exact mass or isotope pattern and can, reciprocally, estimate these from a formula. 
\Rpackage{OrgMassSpecR} \cite{orgmassspecr2012} has similar capabilities including specific functions to process peptide and protein data: it allows the user, for example, to digest proteins, fragment peptides and estimate peptide isotopic distributions modified peptides with, for example, variable $^{15}N$ incorporation rates. In the \Rpackage{RforProteomics} documentation, we demonstrate how to assess protein abundance of the yeast enolase spike present across the 6 \texttt{PXD000001} channels using \Rpackage{OrgMassSpecR}'s \Rfunction{Digest} function and observe that, allowing for one missed cleavage, we observe 13 out of 79 peptides with length greater than 7 residues (corresponding to the shortest identified ENO peptide), as illustrated in figure \ref{fig:seq}. The \LaTeX~code producing the alignment for the figure has been generated automatically, from within \texttt{R}, using the protein sequence and observed peptide sequences and \TeXshade~\cite{texshade2000}. 

\begin{figure}[!htb]
  \begin{center}
    \begin{texshade}{Figures/P00924.fasta}      
      \setsize{numbering}{footnotesize}     
      \setsize{residues}{footnotesize}     
      \residuesperline*{70}
      \shadingmode{functional}
      \hideconsensus      
      \vsepspace{1mm}
      \hidenames
      \noblockskip
      \shaderegion{1}{16..28}{White}{Blue}    
      \shaderegion{1}{33..50}{White}{RoyalBlue}   
      \shaderegion{1}{61..67}{White}{Blue}   
      \shaderegion{1}{68..79}{White}{RoyalBlue}   
      \shaderegion{1}{89..103}{White}{Blue}   
      \shaderegion{1}{133..139}{White}{RoyalBlue}
      \shaderegion{1}{244..255}{White}{Blue}
      \shaderegion{1}{313..329}{White}{RoyalBlue} 
      \shaderegion{1}{339..346}{White}{Blue} 
      \shaderegion{1}{347..358}{White}{RoyalBlue}
      \shaderegion{1}{376..392}{White}{Blue}
      \shaderegion{1}{416..436}{White}{RoyalBlue}
      \shaderegion{1}{416..437}{White}{Blue}      
    \end{texshade}
    \caption{Visualising observed peptides for the yeast enolase protein. Consecutive peptides are shaded in different colours. The last peptide is a miscleavage and overlaps with \texttt{IEEELGDNAVFAGENFHHGDK}.}
    \label{fig:seq}
  \end{center}
\end{figure}


\subsection{Quality control}\label{sec:qc}

Data quality is a concern in any experimental science, but the high throughput nature of modern \textit{omics} technologies, including proteomics \cite{Beasley-Green2012, Ma2012}, requires the development of specific data exploration techniques to highlight specific patterns in data. Examination of complex data is greatly facilitated by well structured containers such as those cited above, that enable direct access to a specific set of values. This, in turn, streamlines the implementation of default and robust pipelines that recurrently query the same data to produce the diagnostic plots and metrics. It is however also often necessary to manually explore data specificity, making the availability of data management facilities even more important.

\begin{figure}[!htb]
  \begin{center}
    \includegraphics[width=0.32\linewidth]{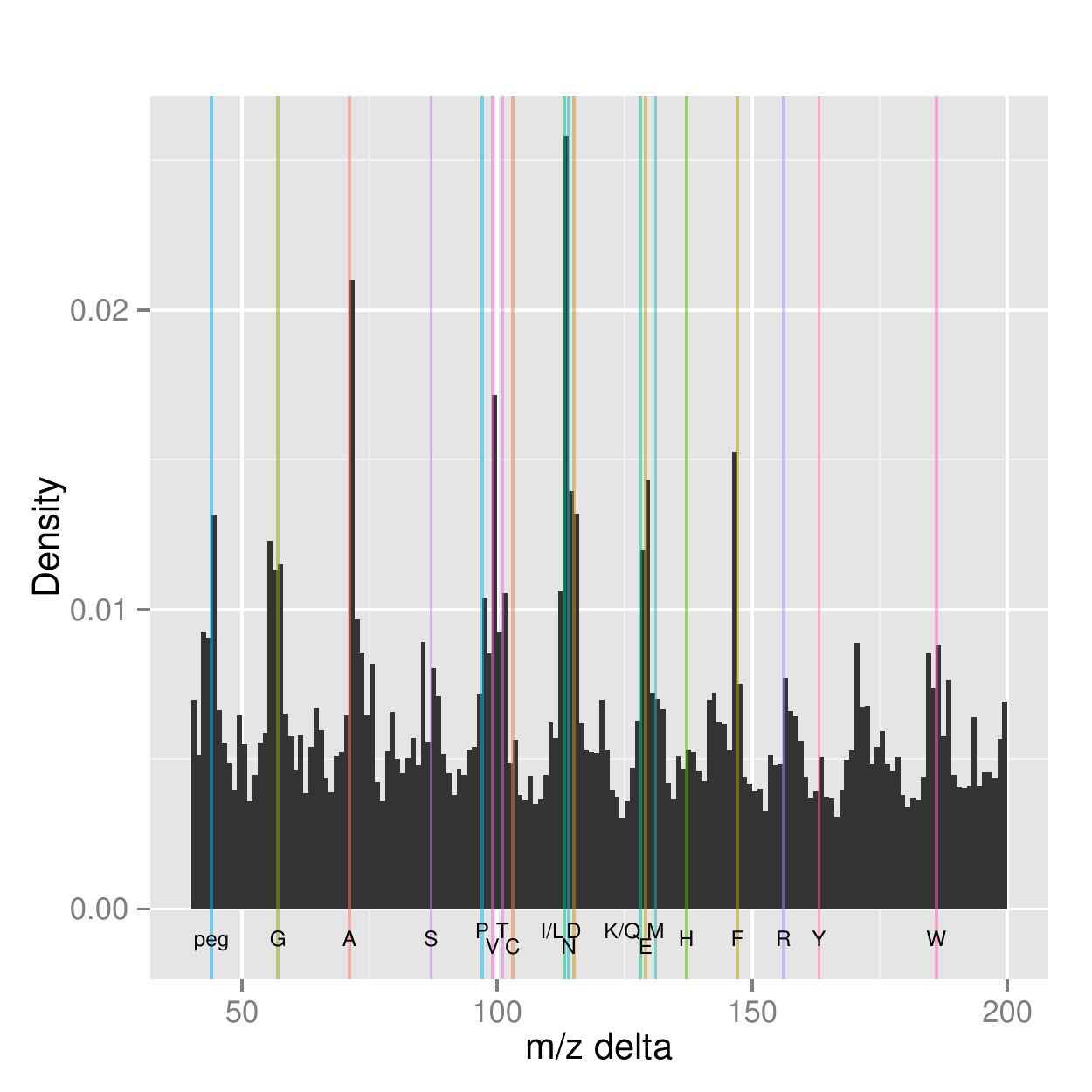}
    \includegraphics[width=0.32\linewidth]{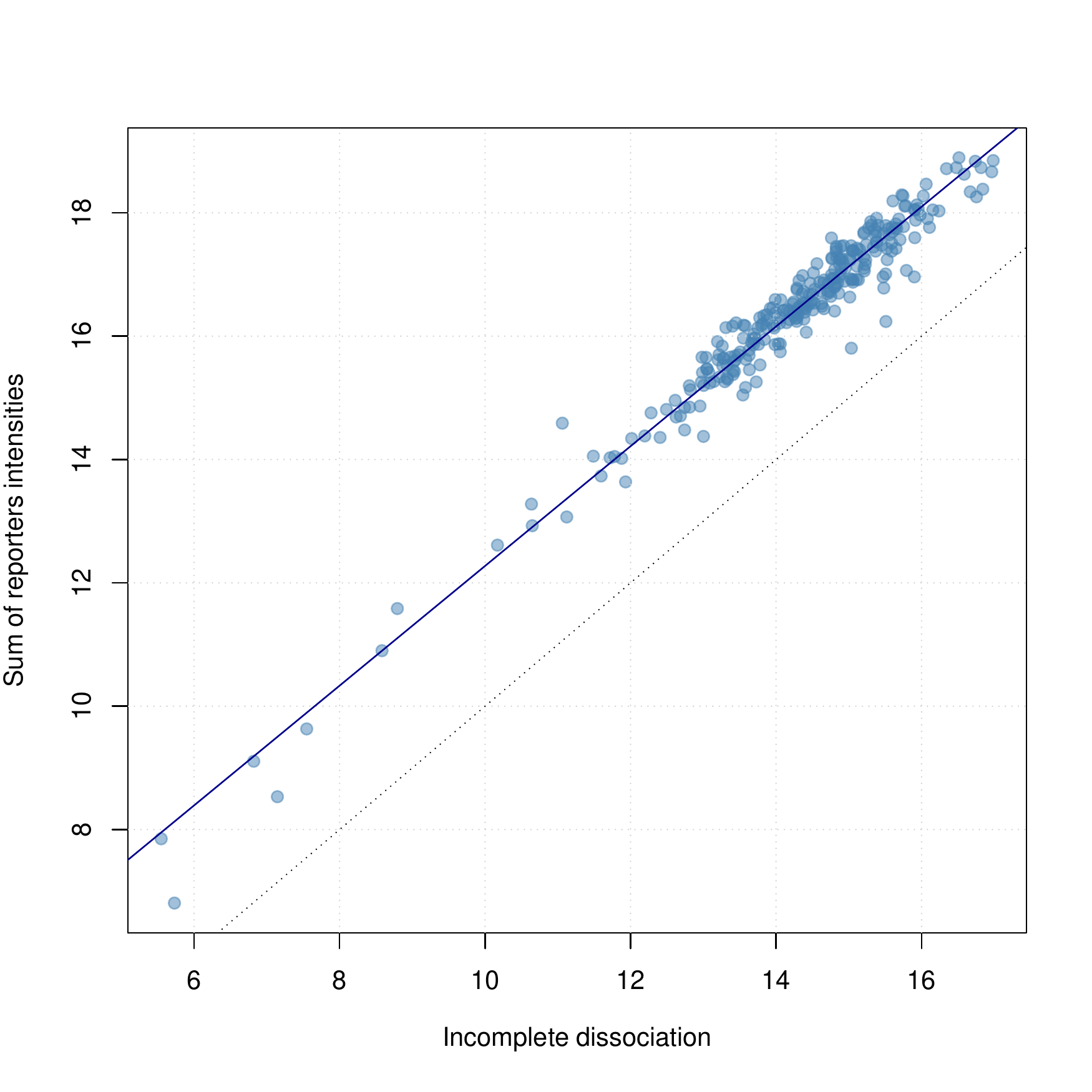}
    \includegraphics[width=0.32\linewidth]{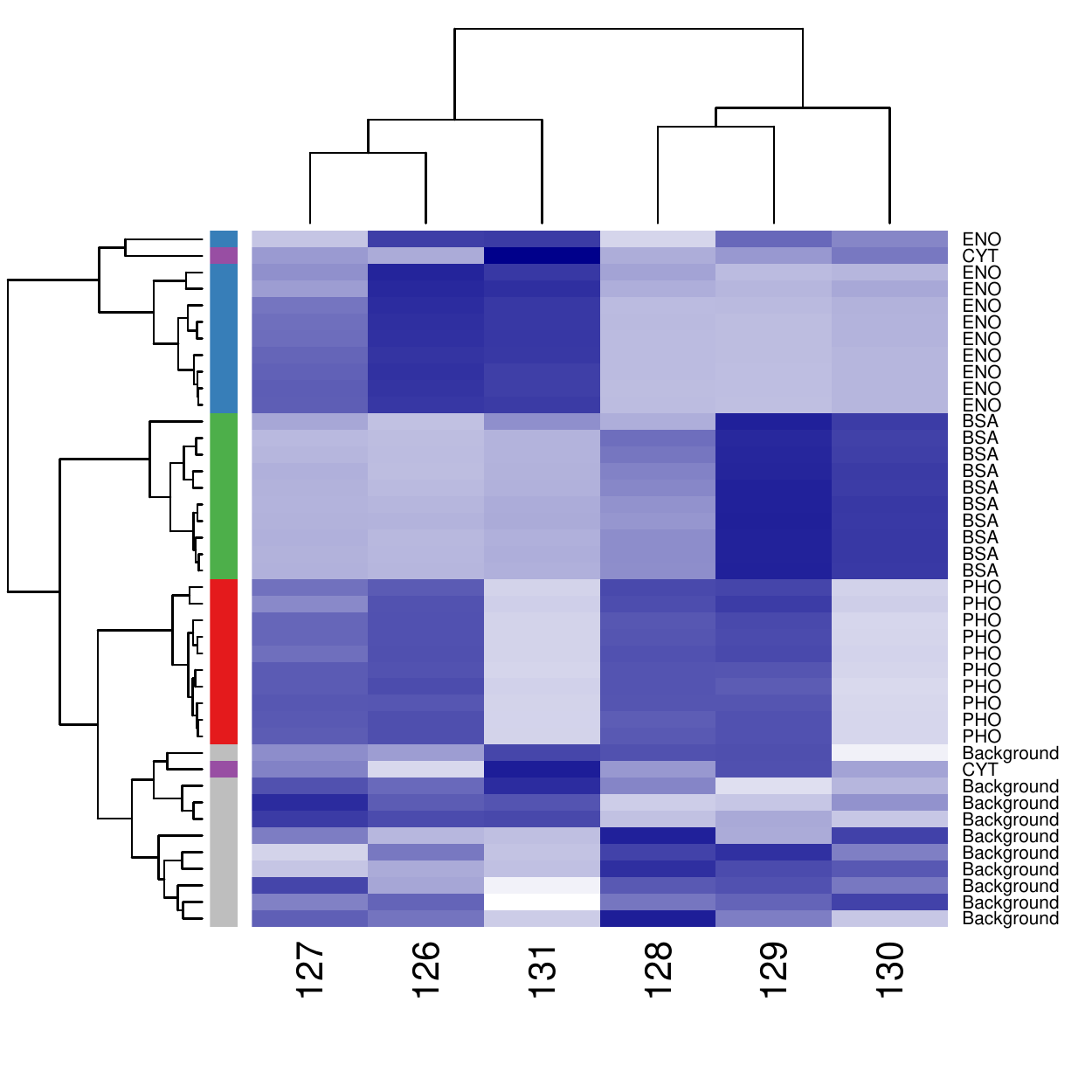}
    \caption{Assessing the quality of the \texttt{PXD000001} data set. On the left, the delta \textit{m/z} plot illustrates the relevance of the raw MS$^2$ spectra for peptide identification. The middle figure compares fully dissociated reporter signal against incompletely dissociated ions, indicating satisfactory reporter dissociation for the experiment. The last figure, a heatmap of a subset of peptides, highlights the expected lack of sample grouping and tight peptides clustering. The first plot is produced by the \Rfunction{plotMzDelta} function from the \Rpackage{MSnbase} package. The other figures used standard base \R plotting functionality. The detailed code and data to reproduce the figures is available in companion package.}
    \label{fig:qcplot}
  \end{center}
\end{figure}

In this section, we present 3 quality plots (figure \ref{fig:qcplot}) that can be used to assess the intrinsic features of the \texttt{PXD000001} data set at different levels. On the left, the distribution of MS$^2$ delta \textit{m/z} \cite{Foster2011} allows the user to assess the relevance of peptide identification; high quality data show \textit{m/z} differences corresponding to amino acid residue masses rising well above the general noise level in the histogram. One can also observe a peak at 44 Da, corresponding to the mass of a polyethylene glycol (PEG) monomer, a common laboratory contaminant in MS.
The middle figure illustrates incomplete dissociation of TMT reporter tags, a technical characteristic of the labelling approach. 
Incomplete dissociation of the reporter and balance moieties of isobaric tags result in this additional single fragment ion peak, in which the multiple channels of quantitation remain convoluted.
The figure illustrates the sum of genuine reporter peaks as a function of incompletely dissociated reporter data. The dotted line corresponds to equal real and lost signal. A linear model has been fitted to the data (blue line), indicating that there is, on average, 100-fold more genuine reporter signal. 
The heatmap on the right indicates the relevance of our quantitation data at the level of our experiment. Congruent peptide clustering indicates agreement between spike peptides while no significant grouping is detected for the samples.

Although the figures above are helpful individually, quality assessment is often most efficient when put into context. Lab-wide monitoring of quality properties and metrics over time to gain experience of average performances and critical thresholds, is the most efficient and valuable application of quality control; the tools presented in this section are one way to automate such a process.

\subsection{Data analysis}\label{sec:analysis}

In this section, we will describe data analysis pipelines for two quantitative strategies, namely MS$^E$label-free and isobaric tagging, using \Rpackage{synapter} and \Rpackage{isobar} respectively. 

Once quantitation data is obtained, it is often desirable to correct technical biases to improve detection of biologically relevant proteins. The availability of well established normalisation algorithms within the Bioconductor project are directly applicable here. The \Robject{MSnSet} object called \Robject{qnt}, created in section \ref{sec:ms2quant} can be normalised using various methods, including quantile normalisation \cite{Bolstad03} and variance stabilisation \cite{Huber2002, Karp2010} using a single \Rfunction{normalize} command. \Rpackage{isobar} also has similar functionality, tailored for \Robject{IBSpectra} objects; its \Rfunction{normalize} method corrects by a factor such that the median intensities in all reporter channels are equal. 

\begin{figure}[!htb]
  \begin{center}
    \includegraphics[width=0.9\linewidth]{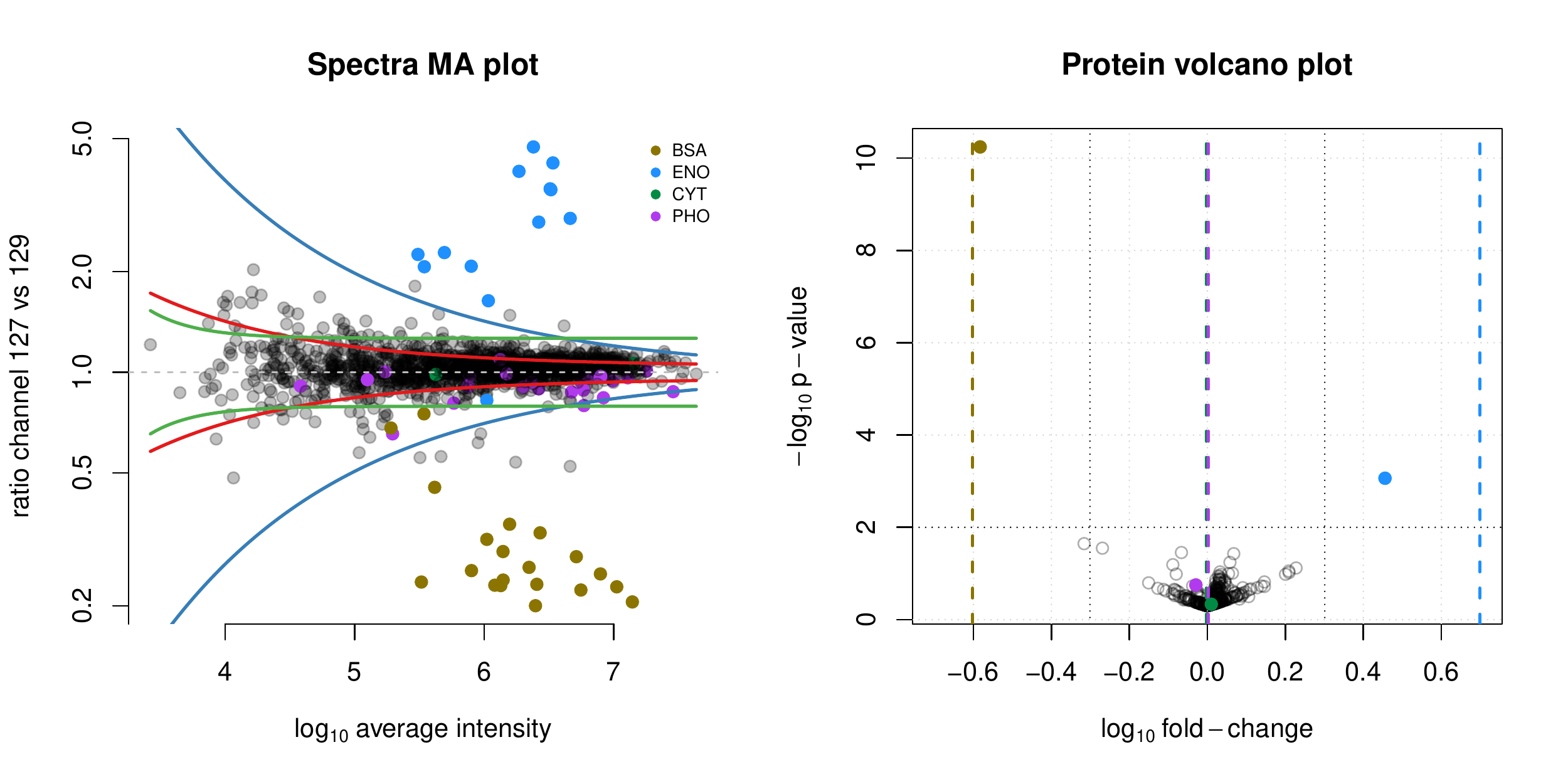}
    \caption{ On the left, the MA plot for the \texttt{PXD000001} 127 vs. 129 reporter ions, showing the 95\% confidence intervals of the background peptides (red), spikes (blue) and all (green) peptide noise models. The respective peptides are colour-coded according to the proteins. The volcano plot on the right illustrates protein significance ($-log_{10}$ p-value) as a function of the $log_{10}$ fold-change. The vertical coloured dashed indicate the expected $log_{10}$ ratios. The black dotted horizontal and vertical lines represent a p-value of 0.01 and fold-changes of 0.5 and 2 respectively. }
    \label{fig:res1} 
  \end{center}
\end{figure}

\Rpackage{isobar} implements methodology to model variability in the data. We will illustrate this using the \texttt{PXD000001} data to estimate spectra and proteins exhibiting significant differences between channel 127 and 129. As shown on figure \ref{fig:res1}, experimental noise has been approximated using the \Rfunction{NoiseModel} function on \textit{Erwinia} background (red), spiked-in (blue) or all (green) peptides (left) and protein ratios and significance have been computed (using the full noise model) with the \Rfunction{estimateRatio} function, to call statistically relevant proteins.

\bigskip

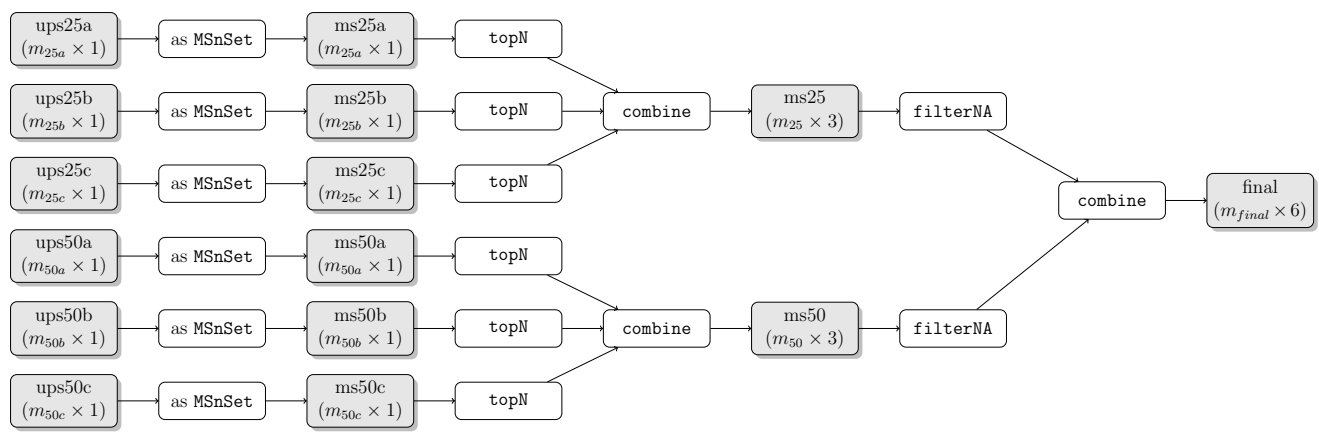
\begin{figure}[!htb] 
  \begin{center}
    \scalebox{0.6}{
      \begin{tikzpicture}[node distance = 3.25cm, auto]
        \node [obj] (ups25a) {ups25a ($m_{25a}\times1$)};   
        \node [fun, right of=ups25a] (as25a) {as \Robject{MSnSet}};
        \node [obj, right of=as25a] (ms25a) {ms25a ($m_{25a}\times1$)};
        \node [fun, right of=ms25a] (topN25a) {\Rfunction{topN}};
        \node [obj, below of=ups25a, node distance=1.6cm] (ups25b) {ups25b ($m_{25b}\times1$)};   
        \node [fun, right of=ups25b] (as25b) {as \Robject{MSnSet}};
        \node [obj, right of=as25b] (ms25b) {ms25b ($m_{25b}\times1$)};
        \node [fun, right of=ms25b] (topN25b) {\Rfunction{topN}};
        \node [fun, right of=topN25b] (comb25) {\Rfunction{combine}};
        \node [obj, right of=comb25] (ms25) {ms25 ($m_{25}\times3$)};
        \node [fun, right of=ms25] (filt25) {\Rfunction{filterNA}};
        \node [obj, below of=ups25b, node distance=1.6cm] (ups25c) {ups25c ($m_{25c}\times1$)};   
        \node [fun, right of=ups25c] (as25c) {as \Robject{MSnSet}};
        \node [obj, right of=as25c] (ms25c) {ms25c ($m_{25c}\times1$)};      
        \node [fun, right of=ms25c] (topN25c) {\Rfunction{topN}};
        \node [obj, below of=ups25c, node distance=1.6cm] (ups50a) {ups50a ($m_{50a}\times1$)};   
        \node [fun, right of=ups50a] (as50a) {as \Robject{MSnSet}};
        \node [obj, right of=as50a] (ms50a) {ms50a ($m_{50a}\times1$)};
        \node [fun, right of=ms50a] (topN50a) {\Rfunction{topN}};
        \node [obj, below of=ups50a, node distance=1.6cm] (ups50b) {ups50b ($m_{50b}\times1$)};   
        \node [fun, right of=ups50b] (as50b) {as \Robject{MSnSet}};
        \node [obj, right of=as50b] (ms50b) {ms50b ($m_{50b}\times1$)};
        \node [fun, right of=ms50b] (topN50b) {\Rfunction{topN}};
        \node [fun, right of=topN50b] (comb50) {\Rfunction{combine}};
        \node [obj, right of=comb50] (ms50) {ms50 ($m_{50}\times3$)};
        \node [fun, right of=ms50] (filt50) {\Rfunction{filterNA}};
        \node [obj, below of=ups50b, node distance=1.6cm] (ups50c) {ups50c ($m_{50c}\times1$)};   
        \node [fun, right of=ups50c] (as50c) {as \Robject{MSnSet}};
        \node [obj, right of=as50c] (ms50c) {ms50c ($m_{50c}\times1$)};
        \node [fun, right of=ms50c] (topN50c) {\Rfunction{topN}};
        \node [fun, below right=1.6cm of filt25] (comb) {\Rfunction{combine}};
        \node [obj, right of=comb] (final) {final  ($m_{final}\times6$)};
        \draw [->] (ups25a) -- (as25a);
        \draw [->] (as25a) -- (ms25a);
        \draw [->] (ms25a) -- (topN25a);      
        \draw [->] (ups25b) -- (as25b);
        \draw [->] (as25b) -- (ms25b);
        \draw [->] (ms25b) -- (topN25b);      
        \draw [->] (ups25c) -- (as25c);
        \draw [->] (as25c) -- (ms25c); 
        \draw [->] (ms25c) -- (topN25c);            
        \draw [->] (ups50a) -- (as50a);
        \draw [->] (as50a) -- (ms50a);
        \draw [->] (ms50a) -- (topN50a);      
        \draw [->] (ups50b) -- (as50b);
        \draw [->] (as50b) -- (ms50b);
        \draw [->] (ms50b) -- (topN50b);      
        \draw [->] (ups50c) -- (as50c);
        \draw [->] (as50c) -- (ms50c);
        \draw [->] (ms50c) -- (topN50c);      
        \draw [->] (topN25a) -- (comb25);      
        \draw [->] (topN25b) -- (comb25);      
        \draw [->] (topN25c) -- (comb25);      
        \draw [->] (topN50a) -- (comb50);      
        \draw [->] (topN50b) -- (comb50);      
        \draw [->] (topN50c) -- (comb50);      
        \draw [->] (comb25) -- (ms25);      
        \draw [->] (comb50) -- (ms50);      
        \draw [->] (ms25) -- (filt25);      
        \draw [->] (ms50) -- (filt50);      
        \draw [->] (filt25) -- (comb);      
        \draw [->] (filt50) -- (comb);      
        \draw [->] (comb) -- (final);              
      \end{tikzpicture}
      }
    \caption{The \Rpackage{synapter} to \Rpackage{MSnbase} pipeline, illustrating how to combine and process data objects in an design specific work flow. Data objects are represented by grey boxes, while functions, that manipulate and transform the objects are shown in white boxes. The respective dimensions of the objects (number of features $\times$ number of sample) are given in parenthesis.}
    \label{fig:synmsn}
  \end{center}
\end{figure}

Data independent MS$^E$ acquisition from a Synapt mass spectrometer (Waters) can be efficiently analysed in \R using the \Rpackage{synapter} pipeline, providing a complete and open work flow (figure \ref{fig:synmsn}) leading to comprehensive data exploration and more reliable results. The test data used for this illustration is a spiked-in set distributed with the \Rpackage{synapterdata} package: 3 replicates (labelled \textit{a} to \textit{c}) of the Universal Proteomics Standard (UPS1, Sigma) 48 protein mix at 25 fmol and 3 replicates at 50 fmol, in a constant \textit{Escherichia coli} background. The set of functions in \Rpackage{synapter} produce data in a specific data container, called \Robject{Synapter} objects, and labelled \Robject{ups} on figure \ref{fig:synmsn}. They store quantitative data for a set of $m$ identified peptides for one unique sample. Although at this step, much has been gained in terms of reliability and number of peptides, we are still far from having interpretable results at this stage. These \Robject{Synapter} objects can easily be converted into \Robject{MSnSet} instances (of dimensions $m_{i} \times 1$, where $m_{i}$ is the number of peptides for the processed sample, labelled \Robject{ms} on figure \ref{fig:synmsn}). Each newly converted MS$^E$ data can now be quantified using the top 3 method \cite{Silva2006} (or any top $n$ variant) where the intensities of the 3 most intense peptides for each protein are aggregated to estimate protein quantities. Each set of replicates is then combined into two new $m_{i} \times 3$ \Robject{MSnSet} instances (named \Robject{ms25} and \Robject{ms50}), one for each set of spike concentration, that are then filtered for missing quantitation, keeping only proteins that have been quantified in at least 2 out of 3 replicates. \Robject{ms25} and \Robject{ms50} are finally combined into the final $m_{i} \times 6$ final data, normalised and subjected to a statistical analysis. 
As illustrated above, it becomes possible to design specific pipelines for any type of experiments using standardised methods and data structures.

\subsection{MS$^2$ spectra identification}

A very recent addition to Bioconductor is the \Rpackage{rTANDEM} package \cite{rTANDEM}.  
The package encapsulates the mass spectrometry identification algorithm X!Tandem \cite{Craig2004}, 
the software for protein identification by tandem mass spectrometry, in \R, 
making it possible to perform MS$^2$ spectra identification within the \R environment 
and directly benefit from \texttt{R}'s data mining capabilities to explore the results. 
The package includes the X!Tandem source code eliminating independent installation of the search engine. 
In its most basic form, the package allows to call the \Rfunction{tandem(input)} function,
where \Robject{input} is either an object of a dedicated class or the path to  a parameter file, 
as one would execute \texttt{tandem.exe /path/to/input.xml} from the command line. 
The results are, as in the original X!Tandem software, stored in an \texttt{xml}, 
which can however be imported into \R in a straightforward way using the 
\Rfunction{GetResultsFromXML} function to subsequently extract the identified peptides 
and inferred proteins. 

\Rpackage{rTANDEM} is currently the only direct \R interface to a search engine and is as such of particularly noteworthy. 
Other alternatives require to execute the spectra identification outside of \R and 
import, export it in an appropriate format and subsequently import is into \R.

\subsection{Annotation infrastructure}\label{sec:annot}

The Bioconductor project provides extensive annotation resources through curated off-line annotation packages, that are updated with every release, or through packages that provide direct on-line access to web-based repositories. The former can be targeted towards specific organisms (e.g. \Rpackage{org.Hs.eg.db} \cite{org.Hs.eg.db2012} for \textit{Homo sapiens}) of systems-level annotation such as gene ontology (the \Rpackage{GO.db} package \cite{GO.db2012} to gain access to the Gene Ontology \cite{Ashburner2000} annotation) or gene pathways (the \Rpackage{reactome.db} \cite{reactome.db2012} interface to the reactome database \cite{Croft2011, DEustachio2011}). \Rpackage{biomaRt} \cite{Durinck2005, Durinck2009} is a very flexible solution to build elaborated web queries to dedicated data mart servers. Both approaches have advantages. While on-line queries allow one to obtain the latest up-to-date information, they rely on network availability and immediate reproducibility in less straightforward to control. 

In the \Rpackage{RforProteomics} documentation, we demonstrate a use case applying 3 complementary alternatives. If one wishes, for example, to extract sub-cellular localisation for a gene of interest, say the human HECW1 gene with Ensembl id \texttt{ENSG00000002746}, it is possible to use (1) the \Rpackage{hpar} package \cite{hpar2012} to query the Human Protein Atlas data \cite{Uhlen2005, Uhlen2010} or (2) to query the \Rpackage{org.Hs.eg.db} and \Rpackage{GO.db} annotations to extract the relevant information or (3) \Rpackage{biomaRt} to query the Ensembl server. Each alternative reports the same location, namely nucleus and cytoplasm, although this might not be necessarily the case. The \Rpackage{hpar} results are very specific and manually annotated, specifying that the protein, although observed in the nucleus, has not been observed in the nucleoli. The other generic alternatives provide additional information, including GO evidence codes.

To conclude this section, we also refer readers to the \Rpackage{rols} package \cite{rols2012}, which provides on-line access to 84 ontologies through the ontology look-up service \cite{Cote06, Cote08}. Among those are the PRIDE, PSI-MS (Mass Spectrometry), PSI-MI (Molecular Interaction) PSI-MOD (Protein Modifications), PSI-PAR (Protein Affinity Reagents) and PRO (Protein Ontology) controlled vocabularies to name those specific to proteomics and mass spectrometry. 

\section{Conclusions}\label{sec:ccl}

We have illustrated data processing and analysis on a set of test and small size data. While real life data sets can be processed on commodity hardware or small servers (see supplementary file of \cite{Gatto2012} and the \Robject{MSnbase-demo} vignette for reports), the sophistication of the biological questions of interest and the increase in throughput of instruments requires software tools to adapt and scale up. \R is an interpreted language (although support for byte code compilation is available through the \Rpackage{compiler} package) and relies in many aspects on a pass-by-value semantics, slowing execution of code compared to compiled languages and pass-by-refence semantics. Fortunately, \texttt{R}'s ability to interoperate with many other languages, including \texttt{C} and \texttt{C++} \cite{Rcpp2011}, allows users to execute computationally demanding tasks while still retaining the flexibility and interactivity of the \R environment. Direct support for parallel computing, large memory/out-of-memory data (see for instance High-Performance Computing task view\footnote{\url{http://cran.r-project.org/web/views/HighPerformanceComputing.html}}) and cloud deployment with the Bioconductor Amazon Machine Image\footnote{\url{http://bioconductor.org/help/bioconductor-cloud-ami/}}, make it possible to embark on large-scale data processing tasks. 

Among the brief list of packages that has been reviewed, we have demonstrated alternative and complementary functionality. 
Most noteworthy however, is the interoperability of these packages, as illustrated in some of the examples. Generally, no specific effort is expected from developers to explicitly promote interaction among packages (on CRAN for example), and thus it is often the user's/programmer's responsibility to implement interoperability.
The Bioconductor project, on the other hand, openly promotes interoperability between packages and
reuse of existing infrastructure. The classes for raw and processed data, briefly described in sections \ref{sec:stdio} and \ref{sec:quant} are adapted from and compatible with existing implementations for transcriptomics data, widely used in many core Bioconductor packages. Data processing procedures used for data normalisation and statistical algorithms are a direct and invaluable side effects of the \R language and previous Bioconductor development. The quality and diversity of available software, fostered by interdisciplinary, open and distributed development, is an immense source of knowledge to build upon.

Although an elaborated environment and programming language like \R has undeniable strengths, its sheer power and flexibility is its Achilles' heel. An important obstacle in the adoption of \R is its command line interface (CLI) that a user needs to apprehend before being able to fully appreciate \texttt{R}. Life scientists very often expect to operate a software through a graphical user interface (GUI), which is probably the major hurdle to the wider adoption of \texttt{R}, or other command line environments, outside the bioinformatics community.
The important point is, however, that properly designed graphical and command-line interfaces are good at different tasks. Flexibility, programmability and reproducibility are the strength of the latter, while interactivity and navigability are the main features of the former and these respective advantages are complementary. Users should not be misguided and adhere to any interface through dogma or ignorance, but choose the best suited tools for any task to tackle the real difficulty, which is the underlying biology.

\bigskip

In this review, we have described how to use \R and a selection of packages to analyse mass spectrometry based proteomics data, ranging from raw data access and visualisation, data processing, labelled and label-free quantitation, quality control and data analysis. It is however essential to underline that, beyond the utilisation of the functionality exposed by the software, fundamental principles of data analysis have been demonstrated.

Every use case that is summarised, including generation of the figures, is  documented in the \Rpackage{RforProteomics} package and is fully \textit{reproducible}: we provide code and data so that interested readers are in a position to repeat the exact same steps and reproduce the same results.
The complexity of biological data itself and the processing it undergoes make it very difficult, even for experienced users, to track the computations and verify the results by merely looking at the input and the output data. As such, \textit{transparency} of the pipeline is a required condition to aim for robustness and validity of the work flow, and the software itself. 
Biology is, by nature, extremely diverse, and creativity in the designs of experiments and the  development and application of technology is the main obstacle to our understanding. The software that is employed must be \textit{flexible} and extensible, to support researchers in their quest rather then limit and constrain them.  Reproducibility, transparency and flexibility are essential characteristics for scientific software, that are provided by the tools described above.

Despite these indisputable advantages, a lot of work still needs to be done to improve and integrate our pipelines, demonstrate how \R can efficiently, reproducibly and robustly be used for in-depth proteomics data comprehension as well as broaden access to these tools to the proteomics community. The \Rpackage{RforProteomics} is one effort in that direction. Finally, support is an essential part of the success and adoption of software; the on-line \R community in general and the the Bioconductor mailing lists\footnote{\url{http://bioconductor.org/help/mailing-list/}} in particular are a rich and broad source of information for new and experienced users. 

\vspace{1cm}

\section*{Acknowledgement} 
The authors are grateful to the \R and Bioconductor communities for providing quality software, robust data analysis methodology and helpful support. This work was supported by the PRIME-XS project, grant agreement number 262067, funded by the European Union 7$^{th}$ Framework Program.


\bibliographystyle{elsarticle-num}
\bibliography{R-for-proteomics}


\end{document}